\documentclass[twocolumn,showpacs,amssymb,aps,nofootinbib,floatfix]{revtex4}
\usepackage{epsfig}

\newcommand{\be}{\begin{eqnarray}}
\newcommand{\ee}{\end{eqnarray}}

\begin{document} \hbadness=10000
\topmargin -0.8cm\oddsidemargin = -0.7cm\evensidemargin = -0.7cm
\title{Why the $a_1$ meson might be a difficult messenger for the restoration of chiral symmetry}
\author{Sascha Vogel, Marcus Bleicher}
\affiliation{Institut f\"ur Theoretische Physik,
J.W.Goethe-Universit\"at, Max-von-Laue-Stra\ss{}e 1, 60438 Frankfurt am Main, Germany}
\date{June 6, 2007}  

\begin{abstract}
We perform a theoretical analysis of the $a_1$ resonance mass spectrum in ultra-relativistic heavy ion collisions within a hadron/string transport approach.
Predictions for the $a_1$ yield and its mass distribution are given for the GSI-FAIR and the critRHIC energy regime. The potential of the $a_1$ meson as a signal for chiral symmetry restoration is explored. In view of the latest discussion, we investigate the decay channel $a_1 \rightarrow \gamma\pi$ in detail and find a strong bias towards low $a_1$ masses. This apparent mass shift of the $a_1$ if observed in the $\gamma\pi$ channel might render a possible mass shift due to chiral symmetry restoration very difficult to disentangle from the decay kinematics. 
\end{abstract}
\maketitle

One of the main goals of relativistic heavy ion physics is to reach densities and temperatures high enough to restore chiral symmetry \cite{Bass:1998vz}. Chiral symmetry is a symmetry of quantum chromodynamics, which is exact if quark masses are zero and approximate if quark masses are small. It is spontaneously broken in nature, but expected to be restored at sufficiently high densities and temperatures. The restoration of chiral symmetry implies a change in the spectral functions of vector mesons (e.g. the $\rho$ meson) and leads to a degeneracy of the spectral functions of the $\rho$ and its chiral partner, the $a_1$ meson. This means that the masses of the chiral partners become equal in the case of full chiral symmetry restoration or approach each other in the case of a partial restoration of the symmetry.
  
Especially the recently observed broadening of the $\rho$ meson spectral function by the NA60 collaboration and the corresponding di-lepton mass spectrum has been interpreted as a signal of chiral symmetry restoration \cite{Koch:1993rj, Brown:1995qt, Cassing:1997jz,Rapp:1999ej, Cassing:1999es}. In fact, the NA60 collaboration measured the $\rho$ meson spectral function in In+In systems at the highest SPS energy of 158~AGeV and observed a deviation from the vacuum Breit-Wigner distribution \cite{Arnaldi:2006jq}. This has triggered various theoretical investigations \cite{Ruppert:2005id,Ruppert:2006hf,vanHees:2006ng}. In summary these studies suggest that some in-medium effects have to be considered, but a conclusive interpretation of the data is still under discussion.
 
 Also the HADES collaboration has recently presented first results on di-electron spectra in light systems at low beam energies (C+C at 2~AGeV) \cite{Agakishiev:2006tg}. Here a deviation from the hadronic vacuum cocktail is visible in the mass region of 500 to 700 MeV. This has been discussed as a possible  observation of partial chiral symmetry restoration and the resulting change in the $\rho$ meson spectral function. Similar data has also been measured by the CERES experiment at CERN in massive nuclear reactions at high energy \cite{Miskowiec:2005dn}.  In spite of the ongoing experimental and theoretical efforts, there are numerous effects that have to be taken into account for a full understanding of data. Thus, it is questionable that a mass shift of the $\rho$ meson alone can be regarded as a ``smoking gun'' signal of chiral symmetry restoration \cite{Vogel:2005pd,Schumacher:2006wc}. 
Therefore a more robust signature of chiral symmetry restoration is needed. Theory predicts that in the case of a full restoration of the symmetry the spectral functions of the $\rho$ meson and its chiral partner the $a_1$ meson become degenerate. The important point is that this statement is independent of any mass shift or broadening. Thus, it has been proposed to measure the $a_1$ mass spectrum in a hot and dense medium and compare it to the mass spectrum of the $\rho$ meson \cite{koch}. If the degeneracy would be observed it is expected to serve as an unambigious experimental signal for the detection  of chiral symmetry restoration in the hot and dense medium.
  
In this paper we argue that the measurement of the $a_1$ (1260) meson may not result in straightforward insights for the understanding and the detection of the chirally restored phase. We discuss the decay kinematics of the $a_1$ meson and argue that an apparent mass shift, or respectively a broadening of the mass spectrum originates from mass dependent branching ratios. 
This effect is not unique to heavy ion reactions, but is nearly independent of energy and system size. 
Furthermore we predict $a_1$ mass spectra for Au+Au and p+p collisions at 20 and 30~AGeV beam energy. The p+p calculations can serve as a vacuum reference.
These systems and energies are experimentally accessible at FAIR-GSI, NA61 and the critRHIC program in the near future.

For our calculations we utilize the UrQMD(v2.3) model, a non-equilibrium transport approach, which relies on the covariant Boltzmann equation. All cross sections are calculated by the
principle of detailed balance and the additive quark model or are fitted to available data. 
UrQMD does not include any explicit in-medium modifications or effects to describe the restoration of chiral symmetry.
The model allows to study the full
space time evolution of all hadrons, resonances and their decay products in hadron-hadron or nucleus-nucleus collisions.
This permits to explore the emission patterns of resonances 
in detail and to gain insight into their origins and decay channels. For previous studies of resonances within this model see \cite{Vogel:2005pd,Bleicher:2002dm,Bleicher:2002rx,Bleicher:2003ij,Vogel:2005qr}.
For further details about the model the
reader is referred to \cite{Bass:1998ca,Bleicher:1999xi}.

Experimentally, the reconstruction of resonances is challenging. One usual technique is to reconstruct the invariant mass spectrum for single events. Then, an invariant mass distribution of mixed events is generated (here, the particle pairs are uncorrelated by definition). This distribution is substracted from the invariant mass spectrum of the single (correlated) events. As a result one obtains the mass distributions and yields (after all experimental corrections) of the resonances by fitting the resulting distribution with a suitable function (usually a Breit-Wigner function peaked around the pole mass of the respective resonance) \cite{Adams:2004ep,Witt:2007xa}.
If a daughter particle (re-)scatters before reaching the detector the signal for the experimental reconstruction is lost. Especially for strongly interacting decay products this effect can be sizeable. In addition, due to the statistical nature of the reconstruction, detailed information on the particle properties and their origin are difficult to obtain. Also possible deviations from a Breit-Wigner distribution can be overseen due to the dependence on the background substraction.\\ 
Thus, we apply a different technique for the extraction of resonances from the model.
We follow the indiviual decay products of each 
decaying resonance (the daughter particles). If the daughter particles
do not rescatter in the further evolution of the system, the resonance is counted
as ``reconstructable''. The advantage of this method is that it allows 
to trace back the origin of each individual resonance to study their spatial and temporal
emission pattern. It also allows to explore the reconstruction efficiency in different decay branches.

The decay channels of the $a_1$ meson are not fully experimentally investigated and details of the branching ratios are unknown \cite{Yao:2006px}. However, a most promising decay channel for the investigation of the restoration of chiral symmetry seems to be the decay $a_1 \rightarrow \gamma \pi$, due to the fact that the photon does basically not interact with the surrounding (hadronic) medium. A study of all other decay channels would imply to study three particle correlations or respectively correlations between resonances and stable particles, which is very tedious, if not impossible in large systems --- however, see also the discussion at the end of this letter.
Thus, experimentally, the $a_1 \rightarrow \gamma \pi$ channel seems the only feasible candidate to measure the $a_1$ meson in heavy ion collisions.

One problem is that the branching ratio into this certain decay channel is still unknown. Further experimental studies in elementary systems are in order to obtain more precise quantitative results in theoretical investigations. While the observed branching ratio $\Gamma_{a_1 \rightarrow \gamma \pi} / \Gamma_{a_1}^{tot}$ $\sim 10^{-3}$ \cite{Yao:2006px} we use an increased branching ratio of $10^{-1}$ in this letter to improve statistics. The shape of the spectra and the discussion of the decay kinematics are essentially independent of the exact value of the branching ratio however one should keep this upscaling in mind for the absolute multiplicities observable in the $\gamma \pi$ channel.

\begin{figure}[htbp]
   \centering
   \includegraphics[width=2.7in]{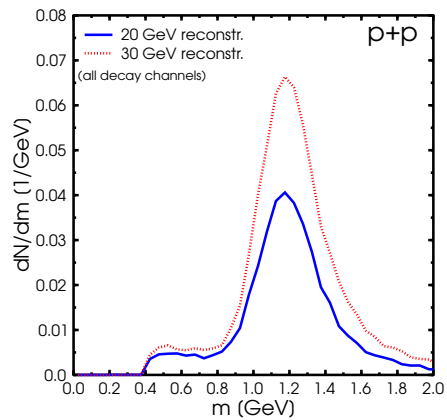} 
\vspace{-.6cm}
   \caption{Mass distribution of $a_1$ mesons in proton-proton collisions at $E_{lab}$ = 20 and 30~AGeV as obtained from UrQMD calculations.}
   \label{mass_spectrum_pp}
\end{figure}

In Fig. \ref{mass_spectrum_pp} the mass spectrum of the $a_1$ meson for p+p collisions at 20 and 30~AGeV obtained from the UrQMD calculation is shown. One observes a clear peak around the pole mass of the $a_1$ meson. At lower masses a shoulder-like structure is visible, which will be of importance in the discussion of the $a_1 \rightarrow \gamma \pi $ channel later on. Note that this mass spectrum is narrower than the one obtained experimentally from $\tau$ decays \cite{Schael:2005am}, see also \cite{Wagner:2007wy} for a detailed discussion of the $a_1$ spectral shape.

\begin{figure}[htbp]
   \centering
   \includegraphics[width=2.7in]{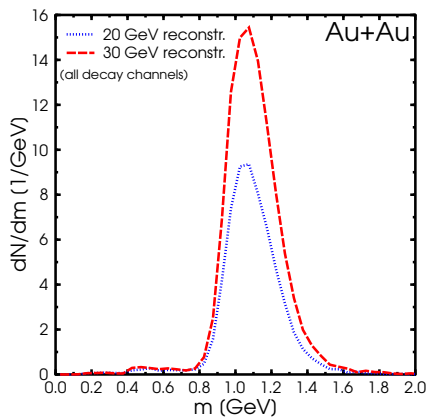} 
\vspace{-.6cm}
   \caption{Mass distribution of $a_1$ mesons in central Au+Au collisions (b $\le$ 3.4 fm) at 20 and 30~AGeV as obtained from UrQMD calculations.}
   \label{mass_spectrum_auau}
\end{figure}

In Fig. \ref{mass_spectrum_auau} the mass spectrum of $a_1$ mesons for central (b $\le$ 3.4 fm) Au+Au collisions at 20 and 30~AGeV as obtained from UrQMD calculations is depicted. As in the p+p case one observes a peak around the pole mass (although slightly shifted to lower masses for kinematic reasons discussed in \cite{Bleicher:2003ij}) and a shoulder-like structure at lower masses of roughly 500~MeV. This shoulder originates from decays into the $\gamma \pi$ channel and will be discussed in the following.

What happens now if one explicitly triggers on the decay channel $a_1 \rightarrow \gamma \pi$ that seems most suitable for the study of the $a_1$ in heavy ion reactions?
As discussed in \cite{Bass:1998ca,Sorge:1995dp} the branching ratios of resonances depends on the mass of the decay products.
The total decay width $\Gamma_{tot}(M)$ of a resonance is defined as the sum of all partial decay widths and depends on the
mass of the excited resonance:
\begin{equation}
\Gamma_{tot}(M)  
\label{gammatot}
       \,=\, \sum  \limits_{br= \{i,j\}}^{N_{br}} \Gamma_{i,j}(M) ,
\end{equation}
where $\Gamma_{i,j}(M)$ is the partial decay width, M is the mass of the resonance and the summation over $N_{br}$ denotes a summation over all possible decay channels.
The partial decay widths $\Gamma_{i,j}(M)$ for the decay into the 
exit channel with particles $i$ and $j$ is given by \cite{Bass:1998ca,Sorge:1995dp}:
\newpage
\begin{eqnarray}
\label{gammapart}
&&\Gamma_{i,j}(M) \\
        &&\,=\, 
       \Gamma^{i,j}_{R} \frac{M_{R}}{M}
        \left( \frac{\langle p_{i,j}(M) \rangle}
                    {\langle p_{i,j}(M_{R}) \rangle} \right)^{2l+1}
         \frac{1.2}{1+ 0.2 
        \left( \frac{\langle p_{i,j}(M) \rangle}
                    {\langle p_{i,j}(M_{R}) \rangle} \right)^{2l} } \nonumber
\end{eqnarray}
here $M_R$ denotes the pole mass of the resonance, $\Gamma^{i,j}_{R}$
its partial decay width into the channel $i$ and $j$ at the pole and
$l$ the decay angular momentum of the exit channel.
${\langle p_{i,j}(M) \rangle}$ denotes the momentum of the decay products in the center of momentum frame.

\begin{figure}[h]
   \centering
   \includegraphics[width=2.7in]{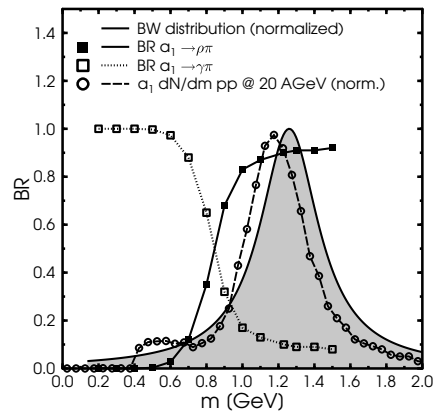}
\vspace{-.6cm}
   \caption{Mass dependent branching ratios for the $a_1$ meson with the two exit channel of $\gamma \pi$ and $\rho \pi$ as calculated from UrQMD. Filled squares depict the branching ratio of $a_1$ mesons into the exit channel $\rho \pi$, whereas open squares depict the branching ratio into $\gamma \pi$. Below a mass of 750~MeV the decay channel $a_1 \rightarrow \rho \pi$ is kinematically suppressed and the channel $a_1 \rightarrow \gamma \pi$ dominates. At masses above 750~MeV the branching ratio into $\rho \pi$ increases steeply. The grey shaded area depicts a normalized Breit-Wigner distribution around the $a_1$ pole mass, whereas the circles depict the normalized mass spectrum of the $a_1$ meson as obtained from UrQMD calculations for p+p collisions at 20~AGeV.}
   \label{massdep_branchingratio}
\end{figure}

Fig. \ref{massdep_branchingratio} shows the branching ratio of the $a_1$ meson as a function of the mass of the $a_1$ as obtained from UrQMD calculations, wherein the definitions \ref{gammatot} and \ref{gammapart} are implemented including the finite width of decay particles. Filled squares depict the branching ratio of $a_1$ mesons into the exit channel $\rho \pi$, whereas open squares depict the branching ratio into $\gamma \pi$. Also shown is a normalized Breit-Wigner distribution (full line) and the normalized mass spectrum of the $a_1$ meson as obtained from proton-proton collisions at 20~AGeV from UrQMD. One observes that at masses lower than 750~MeV the decay channel of 
$a_1 \rightarrow \gamma \pi $ dominates because the decay channel into $\rho \pi$ is kinematically suppressed. At masses greater than 750~MeV the $\rho \pi$ decay channel is dominantly populated and the contribution from the $\gamma\pi$ channel becomes less important. Depicted in Fig. \ref{massdep_branchingratio} are only two of the possible decay channels listed in \cite{Yao:2006px}. All other exit channels consist of even heavier decay products.

\begin{figure}[h]
   \centering
   \includegraphics[width=2.7in]{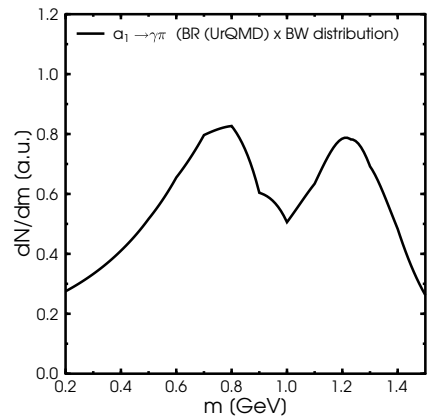} 
\vspace{-.6cm}
   \caption{Sketch of the mass spectrum for the $a_1$ meson as obtained from folding the Breit-Wigner distribution with the branching ratios as shown in Fig. 3. One observes a double peak structure with one peak centered around the nominal pole mass of 1260~MeV and another peak at lower masses.}
   \label{a1_bw}
\end{figure}

Fig. \ref{a1_bw} depicts the mass spectrum (with arbitrary normalization) as obtained from folding the Breit-Wigner distribution of the $a_1$ meson with the mass dependent branching ratio as depicted in Fig. \ref{massdep_branchingratio}. One observes a double peaked structure, with one peak at the nominal pole mass of 1260~MeV and an additional peak at much lower masses. Also note that it is not completely independent of the partial decay widths, i.e. the branching ratios. In this particular case a total width of 400~MeV was assumed with the (enhanced) partial widths of 40~MeV ($a_1 \rightarrow \gamma \pi$) and 360~MeV ($a_1 \rightarrow \rho \pi$).

\begin{figure}[htbp]
   \centering
   \includegraphics[width=2.7in]{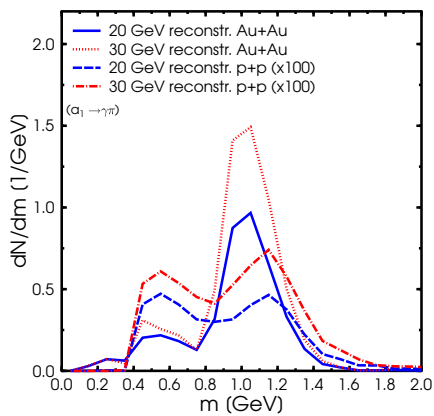} 
   \caption{Mass distribution of $a_1$ mesons which can be reconstructed in  $\gamma \pi$ correlations in nucleus-nucleus and proton-proton collisions at 20 and 30~AGeV. Note that the p+p curves have been multiplied by 100 for visibility, and all calculations use the enhanced branching ratio into $\gamma \pi$. }
   \label{mass_spectrum_gpi}
\end{figure}

After these semi-quantitative discussions, it is clear that a non-trivial $a_1$ mass spectrum has to be expected in the full UrQMD calculation, if a trigger on the $\gamma\pi$ exit channel is employed.
Let us therefore check this effect within the full transport model calculation.
Fig. \ref{mass_spectrum_gpi} depicts the mass spectrum for those $a_1$ mesons which can be reconstructed in the $a_1 \rightarrow \gamma \pi$ decay channel. One observes a clear double peak structure, with one peak centered around the pole mass and one peak in the range of 400-600~MeV. The low mass peak is even more prominent in p+p than in Au+Au collisions.  
Thus, a possible $a_1$ mass shift due to chiral symmetry restoration might be difficult to distinguish from a scenario without mass shift but including mass dependent branching ratios.


Measuring the $a_1$ from a correlations of $\rho$ mesons and pions might be more robust, however is experimentally even more demanding. The $\rho \pi$ channel will avoid the problem of mass dependent branching ratios. Indeed, the HADES (and later the CBM) experiments offer the unique possibility to measure correlations between $\rho$ mesons and pions, where the $\rho$ mesons are reconstructed via the decay channel $\rho^0 \rightarrow e^+e-$. 

These measurements might indeed provide a novel and up to now unexplored route to obtain insights into the transition from the chirally broken to the chirally restored phase.


In summary, we have presented $a_1$ mass distributions for p+p and Au+Au reactions at $E_{lab}$ = 20 and 30~AGeV. It was shown that the reconstruction of the $a_1$ mass distribution in the $\gamma\pi$ channel is strongly biased towards low masses. This can be traced back to the strong mass depence of the $a_1$ branching ratios. Thus, we conclude that a possible observation --- e.g. at the planned GSI-FAIR facility or in a future critRHIC run --- of a modified $a_1$ mass distribution measured in the $a_1 \rightarrow \gamma \pi$ exit channel cannot unambigiously signal an approach towards chiral symmetry restoration. However a measurement in the $a_1 \rightarrow \rho\pi$ channel as it is possible with the HADES and CBM experiments might lead to new information about the approach to chiral symmetry restoration.

The computational resources have been provided by 
the Center for Scientific Computing, CSC at Frankfurt 
University. We thank Volker Koch for fruitful discussions. This work has been financially supported by GSI and BMBF. S.V. thanks the Helmholtz foundation for additional financial support.

\end{document}